# The Strength of Mechanically-Exfoliated Monolayer Graphene

**Xin Zhao[1,2], Dimitrios G. Papageorgiou[1]\*, Liyan Zhu[3,5], Feng Ding[4,5] and Robert J Young [1,4]\***

[1] National Graphene Institute and School of Materials, University of Manchester, Oxford Road, Manchester M13 9PL, UK
[2] Shenzhen Institute of Advanced Graphene Application and Technology, BTR Industrial Park, Xitian, Gongming, Guangming New District, Shenzhen, China
[3] Department of Physics, Jiangsu Key Laboratory for Chemistry of Low-Dimensional Materials, and Jiangsu Key Laboratory of Modern Measurement Technology and Intelligent Systems, Huaiyin Normal University, Jiangsu, China
[4] Institute of Textiles and Clothing, Hong Kong Polytechnic University, Hung Hom, Hong Kong
[5] Center for Multidimensional Carbon Materials, Institute for Basic Science (IBS-CMCM)/School of Material Science and Engineering, Ulsan National Institute of Science and Technology (UNIST), Ulsan 44919, Korea

*E-mail: robert.young@manchester.ac.uk, dimitrios.papageorgiou@manchester.ac.uk

## Abstract

The deformation and fracture behaviour of one-atom-thick mechanically exfoliated graphene has been studied in detail. Monolayer graphene flakes with different lengths, widths and shapes were successfully prepared by mechanical exfoliation and deposited onto poly(methyl methacrylate) (PMMA) beams. The fracture behaviour of the monolayer graphene was followed by deforming the PMMA beams. Through in-situ Raman mapping at different strain levels, the distributions of strain over the graphene flakes were determined from the shift of the graphene Raman 2D band. The failure mechanisms of the exfoliated graphene were either by flake fracture or failure of the graphene/polymer interface. The fracture of the flakes was observed from the formation of cracks identified from the appearance of lines of zero strain in the strain contour maps. It was found that the strength of the monolayer graphene flakes decreased with increasing flake width. The strength dropped to less than ~10 GPa for large flakes, much lower than the reported value of 130 GPa for monolayer graphene, thought to be due to the presence of defects. It is shown that a pair of topological defects in monolayer graphene will form a pseudo crack and the effect of such defects upon the strength of monolayer graphene has been modelled using molecular mechanical simulations.

## 1. Introduction

Since single-layer graphene was firstly isolated from highly-oriented pyrolytic graphite in Manchester University in 2004, it has aroused worldwide attention in materials science and condensed matter physics [1] since it has many excellent properties. It is the strongest material ever measured with very high stiffness and strength [2]. The thermal conductivity of graphene can reach 5000 W/mK [3] and it also has a remarkable electrical conductivity of up to 6000 S/cm [4]. The charge carriers move within monolayer graphene with little scattering under ambient conditions [5]. Its high surface area can theoretically be up to 2630 $m^2$/g and it shows complete gas impermeability [6]. These properties have excited scientist worldwide and created massive expectations for industry to use graphene. Graphene has great potential to be applied to fabricate polymer nanocomposites [7] with good mechanical [8], thermal, electrical [9] and gas barrier properties. The first graphene-based commercial product was the tennis racquet made using graphene nanoplatelets [10]. In early 2017, the world's first graphene mechanical watch was launched with the weight of only 40 g [11].

Graphene is strong and flexible. Its outstanding mechanical properties arise from its $sp^2$-hybridised carbon bonds inside the 2-D graphene honeycomb lattice. These valence bonds inhibit changes in bond angle and length and give rise to very high levels of stored energy during straining. Lee *et al*. first measured the elastic properties and strength of free-standing graphene by nanoindentation and found the Young's modulus of graphene was about 1 TPa and the tensile strength was about 130 GPa [2], although it is thought that this tensile strength value of graphene might be overestimated [12]. For perfect, defect-free graphene, a value of about 100 GPa can be assumed [13]. The behaviour of graphene under extreme dynamic conditions has been studied by Lee *et al*. using a miniaturized ballistic tests and they found that the specific penetration energy for multilayer graphene is 10 times higher than that for macroscopic steel sheets and they also found the radial cracks approximately followed the crystallographic directions after impact [14]. These interesting findings upon the crystallographic orientation of cracks were also confirmed by Kim *et al*. using transmission electron microscopy (TEM) [15].

There are many examples in the literature of optimistic claims of the strength of graphene. The original report of Lee *et al*. [2] stated that their "experiments establish graphene as the strongest material ever measured". In their Nobel Prize lectures Geim [16] acknowledged that "graphene exhibits record stiffness and mechanical strength" and Novoselov [17] spoke of "its unprecedented mechanical strength". Paton *et al*. [18] stated that "graphene is the strongest material known to man with a breaking strength of 130 GPa". With time, however, the claims have become more exaggerated and there are numerous reports, particularly in the online media, of unsubstantiated claims such as through the incorporation of graphene "unbreakable rubber bands that are 200 times stronger than steel are coming soon" [19].

In reality, graphene is invariably produced with a number of intrinsic and physically-introduced defects during preparation. Even mechanically-cleaved graphene prepared from a graphite single crystal, contains many native defects [20]. The defects inside the flakes may dominate the fracture behaviour and control the final fracture strength. In natural graphene there are many different kinds of natural in-plane point defects and edge line defects [20] that might be capable of causing a dramatic drop in the strength. Vlassiouk *et al*. recently found the contribution of graphene to the strength of the resulting nanocomposites was estimated to be only about

10 GPa [21], much lower than the reported value of 130 GPa [2].

In the experiment of Lee *et al*. [2], mechanically-exfoliated monolayer graphene was placed over an array of circular holes of between 1 μm and 1.5 μm in diameter. Nanoindentation was carried out on a very small area of the graphene film over the hole. During the nanoindentation, the deformation was concentrated on a 300 $nm^2$ region or $1.1 \times 10^4$ atoms under the indenter. Within this small region, graphene showed an unprecedented strength due to the low possibility of it containing defects. When the size of graphene becomes larger, however, the situation will be quite different since as the size of the graphene flake increases, the probability of them containing natural defects becomes higher. In this work, the graphene monolayers that were uniformly deformed, had an area more than six orders of magnitude larger (>300 $μm^2$) than the test area in the experiment of Lee *et al*. [2]; the strength of the flakes that fractured during testing was therefore expected to be significantly lower than 130 GPa, as a result of the presence of different types of intrinsic defects. Shekhawat and Ritchie [22] undertook simulation of the effect of line and point defects upon the toughness and strength of nanocrystalline graphene, such as that produced by chemical vapor deposition (CVD), and showed that their presence led to a significant reduction in strength.

In the application of graphene in nanocomposites, mechanically-exfoliated rather than CVD graphene is normally used and a large flake size is of vital importance to achieve good mechanical reinforcement [23, 24]. It is thought that the lateral dimensions of the flakes should be at least 3 μm to achieve good stress transfer from the matrix to the reinforcement [25]. However, when the size of graphene flakes increases, the probability of them containing defects that can cause easy fracture during deformation also increases, which may compromise the mechanical properties of graphene nanocomposites. The aim of this present study is therefore to investigate the fundamental deformation of one-atom-thick monolayer graphene and to understand what level of strength might be expected for flakes of graphene with dimensions large enough to give significant levels of mechanical reinforcement in composites.

## 2. Materials and Methods

The raw graphite single flakes used to prepare the mechanically-cleaved monolayer graphene were purchased from NGS Trading & Consulting GmbH, Germany. The thermal release tape used to exfoliate graphite into graphene was supplied by Nitto Denko Corporation, Japan. PMMA plates were purchased from Panel Graphic Limited, UK and were subsequently cut into small beams of 70 mm in length and 20 mm in width, followed by polishing at the cut edges to minimize scratches and defects. This procedure allows the PMMA beams to be deformed to high strains (~2%) without fracture.

## 2.1 Preparation of exfoliated monolayer graphene on PMMA

For the deformation and fracture studies of monolayer graphene, the monolayer graphene flakes were prepared by mechanical cleavage of single flake graphite. The graphite flake was initially placed in the middle of the Nitto tape and repeatedly peeled with another layer of tape. At the end of this procedure, the material that remained on the tape was a mixture of multilayer graphene flakes of different thicknesses and lateral dimensions. By repeated peeling, the multilayer graphene was finally cleaved into thin graphene sheets. The tape covered with different layers of graphene was subsequently pressed onto the PMMA beam. As a result, graphene flakes with different thickness and lateral dimensions were obtained on the PMMA beam [1].

A Nikon Eclipse LV100ND optical microscope with Nikon cameras DS-Ri2 and DS-Qi2 was used to identify the mechanically-cleaved monolayer graphene on the PMMA. 50× and 100× lens were used to capture high resolution images. Some images were also obtained with a 100× lens in the special colour palette mode in Horiba Raman system, in which the contrast of the nearly-transparent graphene flakes could be greatly enhanced.

## 2.2 Raman spectroscopy

A Horiba LabRAM HR Evolution Raman spectroscopy equipped with blue Argon laser of 488 nm ($E_{laser}$ = 2.53 eV) was utilized to perform the mapping of the graphene monolayer flakes, prepared by mechanical cleavage, on the PMMA beams at different strain levels. Deformation of the monolayer graphene was undertaken by *in-situ* four-point bending of a PMMA beam during which the graphene flake on the top surface was elongated [25]. The strain on the top surface of PMMA was monitored by using a strain gauge. At each strain level, Raman mapping was conducted to create the strain map on the graphene flakes on the PMMA beam. The laser beam was focused on the graphene flakes with 5% laser power (about 1 mW) and a 100× objective lens that produces a laser spot of about 1 μm in diameter. For the band shift rate measurements, the exposure time was 20 s, 3 accumulations were selected for each measurement, with a grating of 600 gr/mm. For the Raman mapping, the exposure time was 10 s, 2 accumulations were selected for each mapping point and the grating used was 600 gr/mm.

For the Raman band shift rate measurements, Raman 2D band spectra were initially obtained from the middle of the graphene flakes. Upon gradually increasing the strain, a Raman spectrum was obtained at each strain level at the same position. After the bending experiment, Raman 2D band positions were plotted as a function of the applied strain and the graphene Raman 2D band shift rates were obtained.

For the whole flake mapping and line mapping procedures, the strain was increased in steps to the strain level at which a fracture on the graphene flake was detected or if no crack was detected, to the maximum achievable strain (~2%). The strain steps were approximately 0.2% for the Raman mapping of the whole flake and 0.1% for the line mapping.

## 2.3 Computational modelling

To show the effect of topological defects on the mechanical behavior of graphene, we simulated the tensile response of a graphene flake with a dimension of around 12 × 24 nm² with two pairs of pentagon-heptagon defects embedded. Periodic boundary conductions were applied along both the *X* and *Y* directions (See Fig. 8a). The unstrained sample was fully relaxed, including ionic positions and lattice constants, by using the conjugate gradient algorithm as implemented in the large-scale atomic/molecular massively parallel simulator (LAMMPS). The interaction between carbon atoms was described by AIREBO typed empirical potential. After structural optimization, we changed the lattice constant along the *Y* direction to mimic the applied tensile load. For each strained structure, we also fully relaxed the whole system to make the residual force acting on each atom less than $1.0 \times 10^{-5}$ eV/nm.

## 3. Results and Discussion

### 3.1 Analysis of graphene deformation

In this study, we investigated the deformation of 10 individual monolayer graphene flakes and a number of graphene micro-ribbons, all with different characteristics. The Raman spectrum of monolayer graphene deposited on PMMA can be seen in **Figure 1a.** In the Raman spectrum for pure PMMA, the characteristic PMMA Raman bands can be seen. The top Raman spectrum, obtained from a monolayer graphene flake deposited on PMMA, shows the characteristic G band of graphene at around 1585 cm⁻¹ and its 2D band at around 2696 cm⁻¹. It should be pointed out that the Raman bands of the graphene are just from an one-atom-thick graphene flake deposited on a 2 mm thick PMMA substrate, indicating the strong resonance Raman scattering from monolayer graphene.

Upon applying uniaxial tensile stress to monolayer graphene flakes, phonon softening occurs with the 2D Raman band undergoing a redshift [26]. **Figure 1b** shows an example of the positions of the Raman 2D band before deformation and at 1.3% and 1.8% strain (a redshift of 110 cm⁻¹). From **Figure 1b** it can also be seen that before deformation the 2D band can be fitted with a single Lorentzian function. With increasing strain, however, the 2D band starts to split and a shoulder appears at higher wavenumber. The reason for this splitting is the distortion of graphene reciprocal lattice and reduced point-group symmetry of the graphene lattice at high strain [27]. In order to fit the Raman 2D band successfully with a single Lorentzian function at high strains, a high energy laser (488 nm) was used to largely supress the inner Raman double resonance process [28]. As a result, with increasing strain, the 2D band splitting initiated only at a relatively high strain (~1.3%) and the Raman 2D band could still be fitted with a single Lorentzian function for strains up to 1.8% (top spectrum in **Figure 1b**).

A typical plot of the Raman 2D band position as a function of strain is presented in **Figure 1c**. Raman measurements were taken from the middle of the flake and a linear shift of the Raman 2D band up to 1.1% strain can be seen, indicating that the interface was still intact after the application of strain at such levels. The slope of the linear fit is -53.6 ± 2.5 cm⁻¹/%, consistent with the monolayer having a Young's modulus of the order of 1 TPa [29]. The slope is slightly lower, however, than the generally accepted Raman 2D band shift rate of around -60 cm⁻¹/% found for free-standing graphene on a substrate under uniaxial tensile strain [25, 30]. After the evaluation of four graphene monolayers, the average redshift of the 2D band was of the order of 55 ± 3 cm⁻¹/%. In the past, similar values of 2D band shifts of -59 cm⁻¹/% [31] and up to -64 cm⁻¹/% [30], have been reported in the literature. These small differences could arise from the variations in strength of van der Waals force between graphene flakes and substrates [31], an uneven stress transfer arising from slippage, the relative movement of the Dirac cones caused by strain [32], the relative crystal orientation of the graphene lattice to the strain direction [33] or finally small variations in laser polarisation direction and laser excitation energy [28]. In order to make a comparison between the fracture behaviour of different exfoliated monolayer graphene flakes, the strain distribution within the graphene flakes needs to be determined using the same Raman 2D band calibration value. In this work, for consistency, the well-accepted 'universal' calibration value of Raman 2D band of -60 cm⁻¹/% for monolayer graphene with Young's modulus of 1050 GPa [25] was used to determine the strains within the monolayer graphene.

During the experimental procedure, strain was applied to the specimens in steps of the order of 0.1-0.2% and strain contour mapping through Raman spectroscopy at each strain level enabled us to monitor the strain distribution in the monolayer graphene flakes. In this way, the strength of the flakes and the modes of failure during deformation could be determined very accurately.

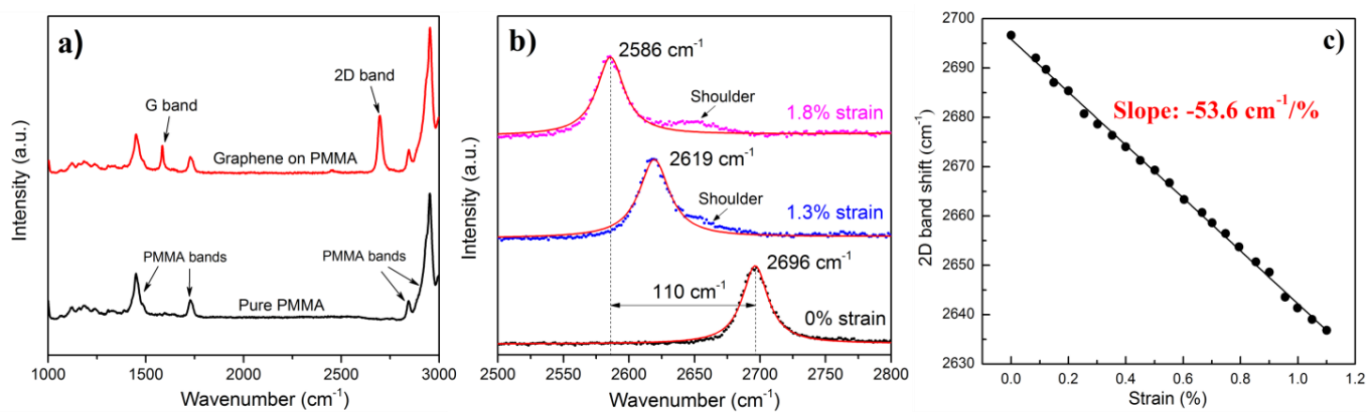

**Figure 1** a) Typical Raman spectra of pure PMMA (bottom) and monolayer graphene on PMMA (top). b) Raman 2D band shift from 0% to 1.8% for a monolayer graphene flake fitted with a single Lorentzian function (obtained using a 488 nm laser). c) Typical shift of the Raman 2D band as a function of strain for a graphene monolayer on PMMA substrate.

### 3.2 Deformation of large flakes

**Figure 2a** shows an optical micrograph of Flake 1 which was 112 μm in length along the strain direction, 18 μm in width at its centre and had an area of 1706 μm². Strain contour mapping was carried out at different strain levels on the flake until a crack was detected. The strain contour map of Flake 1 acquired at 0% strain, before deformation, is presented in **Figure 2b**. It can be seen that the strain distribution over the entire flake is approximately 0% strain. The strain contour mapping at 0.7% strain (before fracture) is shown in **Figure 2c**. In this case, the strain in the middle part is distributed uniformly, with only a small area displaying a strain slightly higher than the strain within the plateau. At the top and bottom edges of the flake, the strain gradually built up over a few microns from 0% to the 0.7% strain plateau along the strain direction. This indicates that stress transfer takes place at the top and bottom edges of the Flake 1 through a shear-lag process [25], arising from the differences in strain between the edges of the flake and the PMMA substrate. Moreover, it should be pointed out that the tape exfoliation technique most commonly induces rough edges that can affect the efficiency of stress transfer in their vincinity. Finally, it should be pointed out that the spatial resolution of the Raman instrument near the edges is in the order of 1 µm, as a result of the laser spot size (~2 µm) and overlapping measurements.

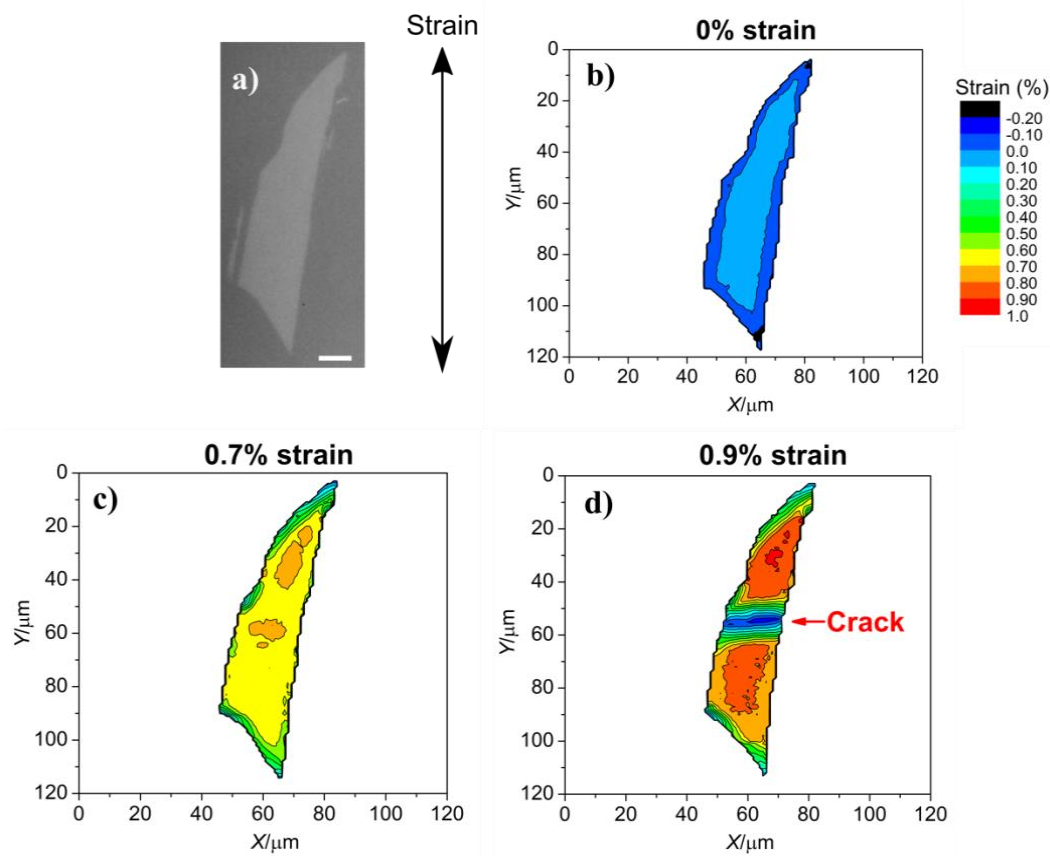

**Figure 2** a) The optical image of the Flake 1 with a scale bar of 10 µm. b) Strain contour map of the flake before deformation (0% strain), c) before fracture at 0.7% strain and d) after fracture at 0.9% strain (mapping step size: 1 µm).

After increasing the strain from 0.7% to 0.9% (**Figure 2d**) a 0% strain region at around 55 µm in *Y* axis can be seen passing through the whole flake, separating the flake into two. This 0% strain region is clearly a crack that formed after loading the graphene monolayer to 0.9% strain, with the flake fracturing at around 0.8% strain. As the Young's modulus of monolayer graphene is 1050 GPa, the tensile strength of Flake 1 can be calculated to be around 8.4 GPa. The strain contour plots in each of the two parts consist of a relatively uniform strain plateau in the middle and well distributed strain field lines at the edges, along the strain direction. This once again indicates that the strain gradually builds up from the top and bottom edges to the middle plateau area (in both parts), consistent with shear-lag theory.

Another large flake (Flake 2) of 84 µm in length and a width of around 30 µm (**Figure 3a**) was subjected to deformation and strain contour mapping was undertaken via Raman spectroscopy. The strain contour plots in monolayer graphene Flake 2 before deformation and after fracture at 1.02% strain, are shown in **Figures 3b** and **3c**. A strain of up to 0.3% was present in the graphene flake before deformation. It represents the residual strain which was generated in this flake by pressing it onto the PMMA beam during specimen preparation. The strain was then gradually increased from 0.3% to 1.02% and during this process two linear mappings were conducted at different strain levels along the L1 and L2 lines shown in the optical image of the monolayer (**Figure 3a**). The strain distributions along the L1 and L2 lines at different strain levels can be seen in **Figures 3d** and **3e**.

From the contour plot in **Figure 3b** at 0.3% strain, it can be seen that the strain is relatively uniform in the middle part of the flake. There are also some low strain areas at the top and bottom areas near the edge that were confirmed by the linear mapping process. During mapping along L1 after the application of strain (**Figure 3d**), there is always an approximately triangular segment of strain distribution appearing at all strain levels, in the *Y* position from 0 to 10 µm, that is similar to the strain distribution along a fibre with length shorter than the critical length [34]. This implies that this area was fractured prior to deformation. Moreover, the wrinkles that can be easily observed at the bottom of the flake in **Figure 3a**, do not transfer stress as effectively as the rest of the flake and especially the middle part, as it can be realised from the strain distribution plot (**Figure 3d**) at *Y* positions between 70 and 84 µm. After gradually increasing the strain from 0.3% to 0.78%, the strain gradually builds up from 0% at 10 µm to a plateau of 0.78% at 20 µm between 10 µm to 84 µm along the *Y* axis. This indicates that stress is transferred effectively between graphene and the PMMA substrate [25]. However, at the bottom part of Flake 2, from 70 µm to 84 µm along the *Y* axis, the strain cannot build up at all strain levels; instead a plateau can always be seen, with strain consistently lower than the middle of the flake, which arises from the irregular distribution of the wrinkles on the bottom area near the edge of the graphene flake (**Figure 3a**), that inhibit the build-up of strain.

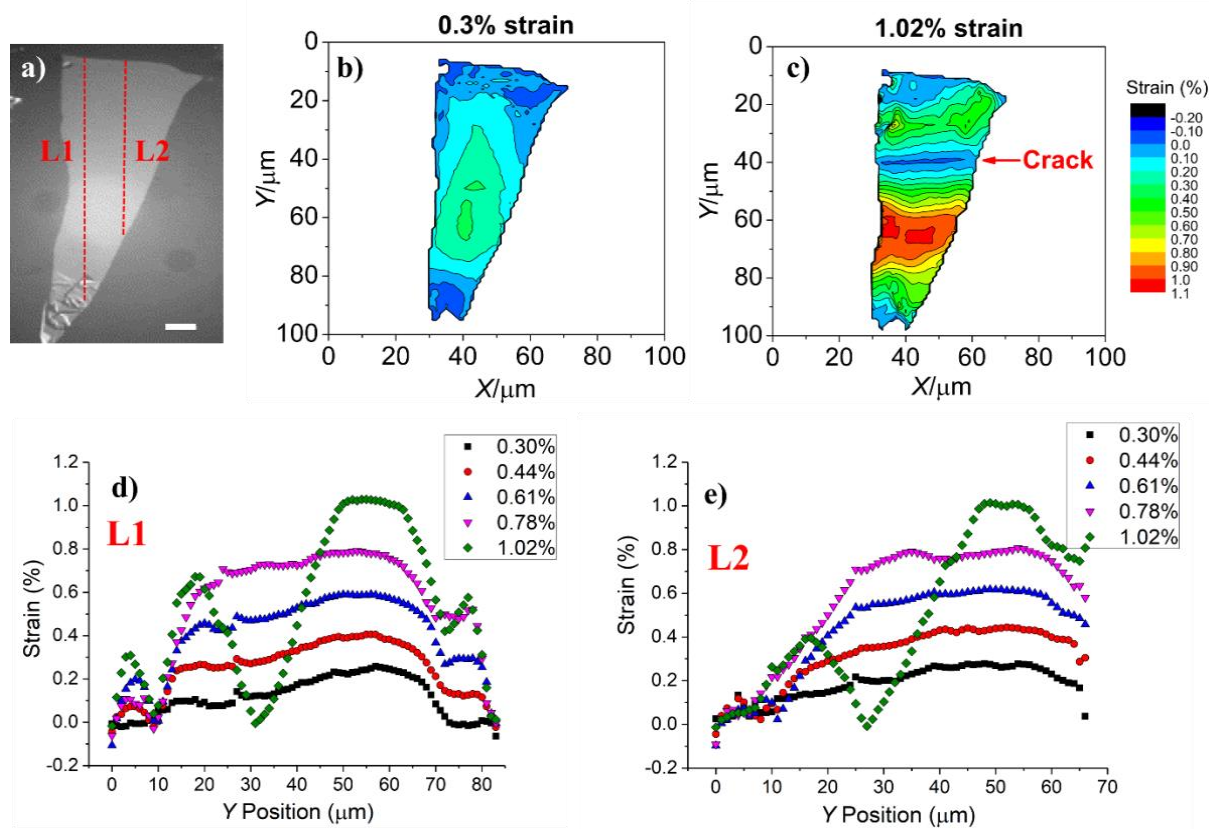

**Figure 3** a) Optical image of Flake 2 with two mapping lines L1 and L2 (linear mapping direction is from top to bottom - scale bar: 10 µm). b) Strain mapping of Flake 2 displaying up to 0.3% residual stress before bending. c) Strain mapping of Flake 2 after fracture at 1.02% strain. d) & e) The strain distributions along lines L1 and L2

When mapping along L2 (**Figure 3e**), from 0 µm to 10 µm in *Y*, at all strain levels, there is always a plateau of around 0% strain, which is caused by the existence of the crack observed during the linear strain mapping along L1 and/or interface failure in this area, where the stress cannot be effectively transferred. In both **Figures 3d** and **3e**, after increasing the strain from 0.78% to 1.02%, the plateau observed in the middle of the flake, dropped to 0% strain at around 30 µm in *Y*. Such a process generated a triangular strain distribution from 10 µm to approximately 30 µm in *Y* position and a trapezoid strain distribution (from 30 to 70 µm). This is similar to the strain distribution along fibres in a composite with lengths shorter and longer than the critical length, respectively. This indicates that the monolayer graphene fragmented into two parts; one with a length shorter than critical length and another with a length longer than the critical length. Therefore, the critical length for this flake is about 40 µm. This crack can be identified by the 0% strain line passing through the middle of the flake in the strain contour map at 1.02% strain in **Figure 3c**. The existence of the crack at 1.02% strain was also confirmed by optical microscopy (see Figure S1 in the Supporting Information). This crack separates Flake 2 into two different parts. At 1.02% strain, it can be observed that the strain field lines are well distributed on the top edge of the bottom part along the strain directions, which suggests that the strain builds up gradually from the edge to the middle strain plateau in the bottom part, consistent with the shear-lag theory [34]. Flake 2 must have fractured at around 0.9% strain, which corresponds to about 9.5 GPa tensile strength, since the Young's modulus of monolayer graphene is 1050 GPa.

It is possible to determine the interfacial shear stress between graphene and the PMMA substrate before and after fracture. Ignoring the two ends with the pre-existing crack and wrinkles, at 0.78% strain (before fracture) the variation of the graphene strain along the L1 line across the middle of the flake can be fitted using shear lag theory [25]:

$$e_f = e_m[1 - \frac{\cosh(ns\frac{x}{l})}{\cosh(\frac{ns}{2})}] \qquad (1)$$

$$\text{where } n = \sqrt{\frac{2G_m}{E_g}[\frac{t}{T}]} \qquad (2)$$

and $e_m$ is the strain in PMMA substrate, $e_f$ is the strain in the graphene flake with position, $x$, $G_m$ is the shear modulus of matrix, $E_g$ is the Young's modulus of graphene, $l$ is length of graphene, $t$ is thickness of graphene, $T$ is total thickness of the resin, $s$ is the aspect ratio of graphene expressed by $l/t$. The parameter $n$ is related to the efficiency of interfacial stress transfer. From **Figure 4a**, it can be observed that the shear-lag theory performs well in modelling the experimentally-determined variation of strain in Flake 2 at an overall strain of 0.78% and a value of $ns = 9.3$ gives the best theoretical fit with the experimental data.

The shear stress at the interface can be calculated by assuming that the shear stress is balanced by the variation of strain in the graphene flake through the equation [25]:

$$\frac{de_f}{dx} = -\frac{\tau_i}{E_g t} \qquad (3)$$

In this case, the shear stress distribution along this region can be predicted theoretically by [25]:

$$\tau_i = nE_g e_m \frac{\sinh(ns\frac{x}{l})}{\cosh(\frac{ns}{2})} \qquad (4)$$

The theoretical shear stress distribution along L1 line is displayed in **Figure 4b**, in which the shear stress equals to 0 MPa in the middle of the flake and reaches the maximum shear stress of $\pm 2.2$ MPa in the two ends.

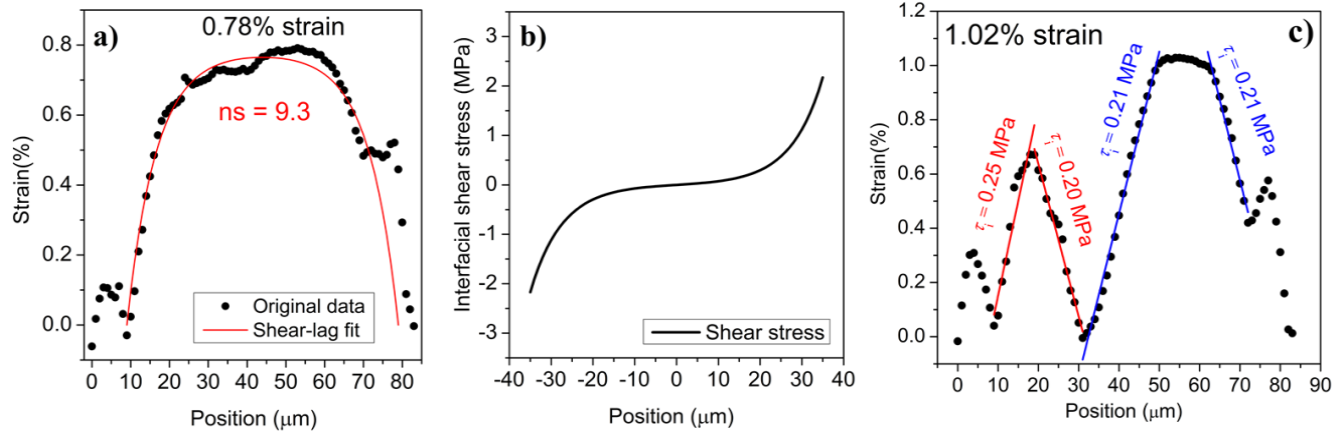

**Figure 4** a) Shear-lag theoretical fit by equation (1) at 0.78% strain along L1 line for Flake 2, b) shear stress distribution predicted by equation (4), c) strain distribution along L1 line at 1.02% strain, when the fracture of Flake 2 took place. The values of $\tau_i$ were determined using equation (3).

At 1.02% strain, fragmentation of the graphene took place; this generated a triangular and a trapezoid strain distribution in the strain distribution graph (**Figure 4c**). The slopes of the linear fitting on both sides of the triangular strain distribution (from 10 to 30 µm) are slightly different. In contrast, in the trapezoid-shaped strain distribution (*Y* position: 30-70 µm), the slopes are very similar. This is an indication that after fracture, the top part of Flake 2 slipped due to imbalanced shear stresses on both sides but the bottom part of Flake 2 did not slip having balanced shear stresses on both sides. The calculated interfacial shear stress after fracture is between 0.20 and 0.25 MPa, which is significantly lower than the maximum shear stress at two ends (2.2 MPa before fracture) (**Figure 4b**). This leads us to the conclusion that the graphene fragmentation process damaged the interfacial adhesion between graphene and PMMA substrate.

### 3.3 Deformation of small flakes

Similar strain contour mapping experiments were also carried out upon some smaller monolayer flakes (Flakes 3, 4) from 0% strain up to the highest achievable strain, in around 0.2% strain steps. The optical micrographs and contour mappings of Flake 3, (45 µm in length along the straining direction and ~12 µm wide in the middle region) and of Flake 4 (60 µm length along the straining direction and ~9 µm wide in the middle region) are shown in **Figure 5**. Here, only the initial mapping at 0% strain and mapping at the highest achievable strain are presented (**Figures 5b, c** and **e, f**).

At 0% strain, for both flakes, the strain is uniformly distributed. Furthermore, at the highest strain (**Figures 5c** and **5f**), except for the low strain areas at the top of the flakes, the strain field lines are also uniformly distributed from 0% strain to the highest strain plateau in the middle of the flake, along the strain direction. The low strain areas at the top of Flakes 3 and 4 can be attributed to cracks formed near the top edges. In the bottom tapering edges in both flakes, the strain did not gradually drop to zero but fell abruptly at the flake end. This situation is therefore similar to a high modulus fibre with a pointed tip, in which the strain drops to zero only very close to the end of the tip [25].

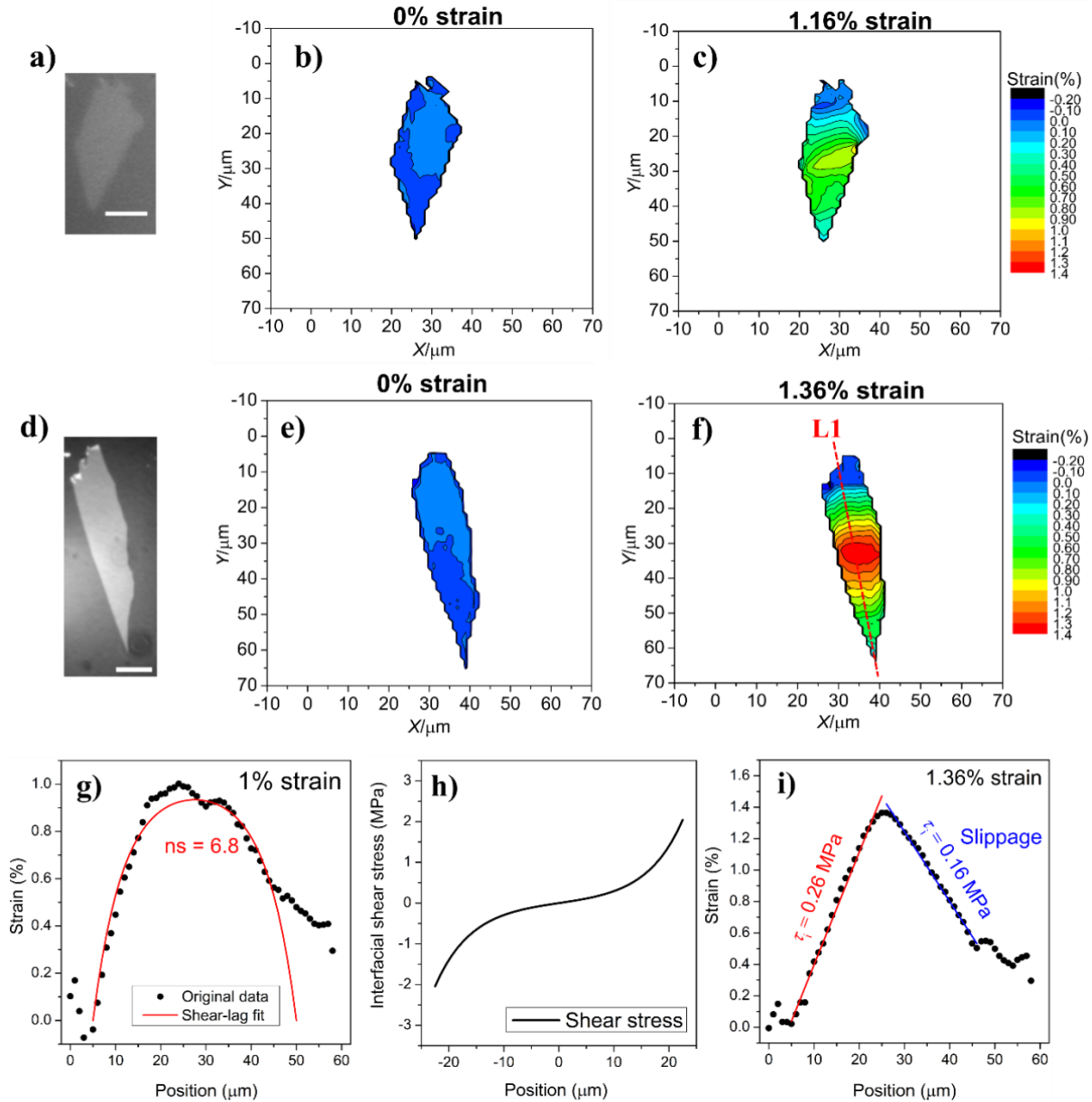

**Figure 5** a) Optical image of Flake 3 with scale bar of 10 µm shown in the bottom right corner. b) & c) Strain contour maps at 0% strain before deformation and at the highest achievable strain for Flake 3. d) Optical image of Flake 4 with scale bar of 10 µm shown in the bottom right corner. e) & f) The strain contour map at 0% strain before deformation and at the highest achievable strain for Flake 4. g) fitting result by shear-lag equation (1) at 1% strain for Flake 4, h) shear stress distribution along Flake 4, i) strain distribution in Flake 4 along L1 line at 1.36% strain after linear fitting by equation (3).

To study the interfacial shear strength before the application of the maximum strain, the strain distribution along the L1 line at 1% strain for Flake 4 (**Figure 5f**), was fitted using equation (1) and the product $ns$ was calculated ($ns = 6.8$) (**Figure 5g**). Based on the $ns$ value, the interfacial shear stress along the L1 line was also calculated and is presented in **Figure 5h**. The maximum interfacial shear stress at the two ends is about ±2.1 MPa.

Furthermore, the strain distribution along line L1, at the maximum achievable strain (1.36% strain) for Flake 4 can be seen in **Figure 5i**. Ignoring the top cracked part and the tapering edge at the bottom, the strain distributions from the two ends to the middle tip can be fitted linearly by equation 3 [25]. A large difference in the slopes can be seen on either side of the triangular strain distributions at the middle of the flake, signifying that the shear stress is not balanced and this caused slippage at 1.36% strain. This slippage prevents the strain from increasing further. Apart from that, Flake 4 is not long enough to support a further increase in the strain along the strain direction. The shear stresses calculated from the slopes of the strain distribution graph are between 0.16 and 0.26 MPa. Based on this result, the maximum interfacial shear stress (2 MPa) is much higher than that after slippage ($\tau_i$ = 0.16 - 0.26 MPa) (**Figure 5g**). This indicates that after slippage, the interface was damaged for Flake 4. At 1.16% strain and 1.36% strain, the strain contour mapping graphs for Flake 3 and 4 show no cracks in the middle part of the flakes. This implies that the tensile failure strain of Flake 3 is higher than 1.16% strain and for Flake 4 is higher than 1.36% strain, which correspond to strengths of >12 GPa and >14 GPa, respectively.

### 3.4 Deformation of micro-ribbons

The deformation of three graphene micro-ribbons was also studied by line mapping along their length. The optical images of the three micro-ribbons (Flakes 5, 6 and 7) are displayed in **Figure 6a**. The lengths and widths for Flakes 5, 6 and 7 were 87 μm, 77 μm and 60 μm and around ~1.5 μm, ~1.0 μm and ~0.8 μm respectively. In **Figures 6b, c** and **d**, the strain distribution plots confirm that the initial strain along the three flakes is close to 0%. With increasing strain, it can be observed that for all micro-ribbons the strain gradually builds up from 0% in both ends and then reaches a plateau in the middle, indicating that the shear-lag analysis can once again explain the interfacial stress transfer between the thin graphene flakes and the PMMA substrate. For Flake 5, the strain was successfully transferred from the PMMA beam to the flake for applied strains up to 1.14%, while a further increase led to a reduction of the strain, as it can be seen from the middle plateau area in **Figure 6b**.

In order to estimate the interfacial shear stress between the flake and PMMA before slippage, the strain distribution along the Flake 5 at 0.92% strain was determined as shown in **Figure 7a** where the experimental data were fitted with equation 1 by using *ns* = 10.6. The interfacial shear stress distribution along Flake 5 was then calculated based on equation 3 (**Figure 7b**). The maximum shear stress in the two ends is ±3.3 MPa, which is significantly higher than the one calculated after slippage. This again proves that after slippage, the interface between Flake 5 and PMMA was damaged. The average maximum shear stress before fracture or slippage of the flakes is around 2.5 ± 0.7 MPa, while after fracture or slippage the average shear stress is around 0.21 ± 0.07 MPa.

By using the simple Kelly-Tyson model [35] (equation 3) to fit the strain distribution linearly from 0% strain at both ends to the strain plateau at 1.14% strain (**Figure 7c**), it can be seen that the values of the slopes are quite different from each other and they correspond to an interfacial shear stress of 0.33 MPa and 0.13 MPa for the top and bottom ends of the Flake 5. This result indicates that the interfacial shear stress at the two ends of Flake 5 was not balanced and this caused the slippage at 1.14% strain. Similar observations were noted for Flakes 6 and 7 at strains higher than 1.06% and 0.85% strain, respectively. The reason for the slippage of these three flakes is that the narrow flakes appear to be stronger than the wider ones, so in our experimental set-up they always debond before fracture. Therefore, on the basis that no fracture was observed for strains up to 1.14%, 1.06% and 0.85% for the three flakes, the calculated strengths for Flakes 5, 6 and 7 are all higher than 12 GPa, 11 GPa and 9 GPa, respectively.

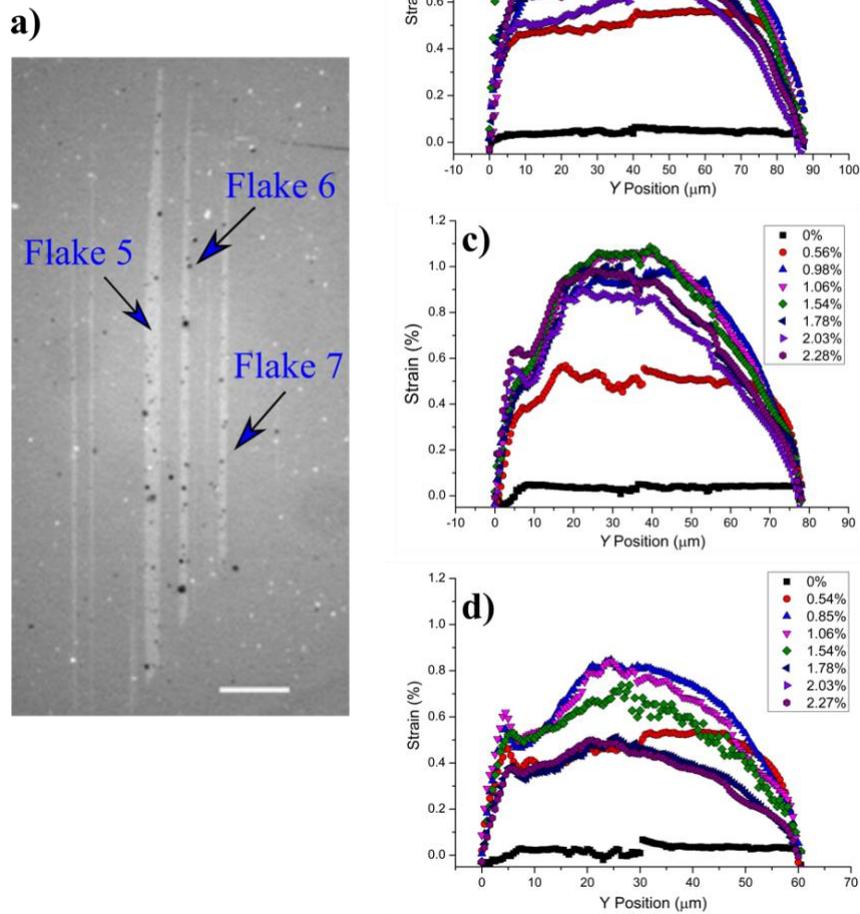

**Figure 6** a) Optical images of Flake 5, Flake 6 and Flake 7 (scale bar: 10 μm), b), c) and d) strain distribution along Flake 5, Flake 6 and Flake 7, respectively.

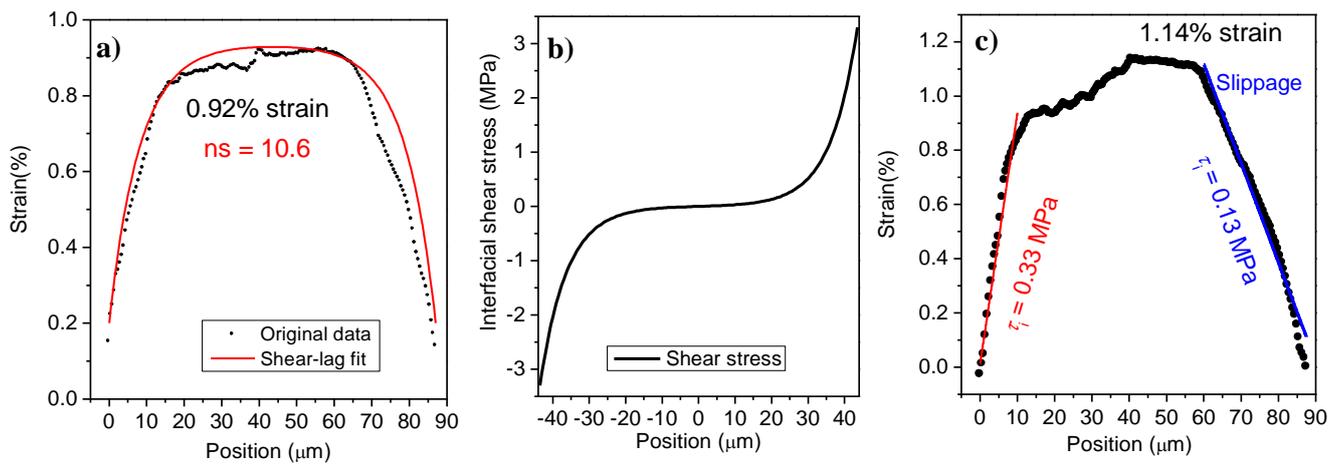

**Figure 7** a) Fitting result by shear-lag equation (3) at 0.92% strain along L1 line for Flake 5, b) Corresponding shear stress distribution predicted by equation (4), c) Strain distribution along L1 line at 1.14% (Flake 5 slipped) with linear fitting by equation (3).

### 3.5 Fracture from internal defects

The deformation and fracture behaviour of a further six different-sized graphene monolayer flakes was also investigated and the results for all 13 flakes are summarised in **Table 1**. It is clear that, in general, narrow flakes are stronger than the wide flakes. This could be due to the probability of finding a defect in the narrow flakes being lower than in the wide flakes. It is well established that the presence of defects in a brittle material can give rise to failure at stress levels well below the theoretical values of strength [36]. The strength is found to decrease as specimen size increases [37]. For example, ~10 μm diameter glass fibers are much stronger than macroscopic specimens of window glass.

The dependence of the strength of a body upon its size is often analysed using the concept of the weakest link in a structure in the statistical analysis of failure, first proposed in the 1950s by Weibull [37]. This is analogous to the breaking of a chain in which failure occurs when the weakest link breaks. When the broken halves are retested they will have a higher strength than that of the original chain. This type of behaviour can be demonstrated for the repeated fracture of glass fibres [37].

The Weibull approach has led to the concept of the probability of failure of a specimen of a brittle material of volume $V$ subject to a stress $\sigma$ being given by an expression of the form [37]

$$P_s(\sigma) = \exp\left[-V\left(\frac{\sigma-\sigma_u}{\sigma_0}\right)^m\right] \tag{5}$$

The parameter $\sigma_u$ is the stress below which fracture is assumed to have zero probability, $\sigma_0$ is a normalising parameter of no physical significance and $m$ is a number, usually termed the Weibull parameter that reflects the variability of the strength. There is a less variable strength for higher values of $m$ and values of $m$ in the range 5-20 are typical for ceramics [37].

This expression can now be used to predict the effect of specimen size upon the strength of material. If we consider two specimens of different volume $V_1$ and $V_2$, it follows that the stresses associated with the same probability of survival, $\sigma_{V1}$ and $\sigma_{V2}$, are given by

$$V_1\sigma_{V1}^m = V_2\sigma_{V2}^m \tag{6}$$

In this analysis it has been assumed that there is no upper limit of flaw size so that $\sigma_u = 0$ [37]. Hence the final expression is

$$\frac{\sigma_{V1}}{\sigma_{V2}} = \left(\frac{V_2}{V_1}\right)^{1/m} \tag{7}$$

which predicts that a larger specimen will have a lower strength.

The monolayer graphene flakes investigated in the study are all of the same thickness. Assuming the flakes that fracture are of similar length and vary mainly in width, $w$, then for the dependence of strength upon flake width equation 7 becomes

$$\frac{\sigma_{w1}}{\sigma_{w2}} = \left(\frac{w_2}{w_1}\right)^{1/m} \tag{8}$$

Hence it is predicted that the strength of a flake should be proportional to the reciprocal of its width to the power $1/m$.

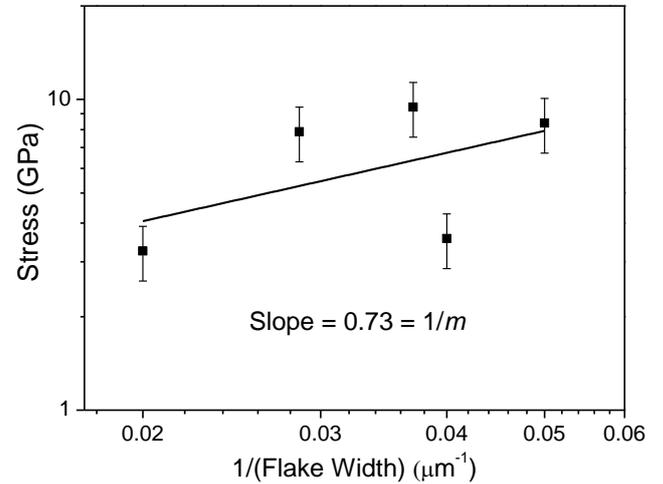

**Figure 8** log-log plot of the strength as a function of the reciprocal of flake width at the crack for the flakes that fractured.

The strength values for the five flakes that fractured (**Table 1**) are plotted as a function of the reciprocal of the flake width in **Figure 8**. The flake width ($X$-axis) for flakes that fractured is the dimension of the flake in the direction perpendicular to the strain axis at the location where the cracks were developed. Although there are only a limited number of points and there is scatter in the data, the slope of the plot is around 0.73 which gives a Weibull parameter $m$ of the order of 1.5. This relatively small value indicates a high variability of strength with size and implies that smaller flakes will be considerably stronger. It is also interesting that the data extrapolate to a strength value of 70 GPa for a flake 1 μm wide and to even higher values for narrower flakes. Shekhawat and Ritchie [22] report a higher Weibull parameter ($m$) of ~11 for polycrystalline graphene. This difference originates from the fact that polycrystalline graphene displays a reduced stress concentration build-up at a crack tip, caused by the more effective distribution of stress at the grain boundaries. Therefore, stress is less concentrated as a larger area of graphene is undergoing deformation. Moreover, Buehler *et al.* have [38] shown that the grain boundaries of graphene cause crack branches and thereby a crack propagates through highly complex branches over the entire material, which leads to greater energy dissipation.

**Table 1.** Characteristics of the flakes (*micro-ribbons) under study and their respective dimensions, strength, maximum interfacial shear stress and failure modes under strain.

| Flake Number | Width (µm) | Length (µm) | Maximum strain (%) | Strength (GPa) | Maximum interfacial shear stress (MPa) | Failure Mode |
|---|---|---|---|---|---|---|
| 1 | 20 | 112 | 0.9 | 8.4 | - | Fracture |
| 2 | 27 | 84 | 1.02 | 9.5 | 2.2 | Fracture |
| 3 | 12 | 45 | >1.16 | >12 | - | Slippage |
| 4* | 9 | 60 | >1.36 | >14 | 2.1 | Slippage |
| 5* | ~1.5 | 87 | >1.14 | >12 | 3.3 | Slippage |
| 6* | ~1.0 | 77 | >1.06 | >11 | - | Slippage |
| 7* | ~0.8 | 60 | >0.85 | >9 | - | Slippage |
| 8 | 50 | 20 | 0.31 | 3.3 | - | Fracture |
| 9 | 15 | 20 | >1.05 | >11 | - | Slippage |
| 10 | 25 | 35 | 0.34 | 3.6 | - | Fracture |
| 11 | 10 | 50 | >2.04 | >21 | - | Slippage |
| 12 | 35 | 140 | 0.75 | 7.9 | - | Fracture |
| 13 | 17 | 150 | 1.84 | >19 | - | Slippage |

A number of different internal crystallographic defects and combinations of defects may be present in graphene monolayer flakes that can subsequently affect the mechanical properties of a flake [20]. It is also possible to undertake analysis of such defects upon the strength of a graphene monolayer using a computational molecular mechanical modelling approach.

Even if the graphene sample is crack free, the mechanical strength can still be reduced dramatically in the presence of topological defects in the sample. To demonstrate this phenomenon, the deformation of a graphene monolayer with dimensions of ~12 × 24 nm² containing two pairs of pentagonal and heptagonal rings [39] into the hexagonal network of graphene (**Figure 9a**) was investigated. It should be pointed out that this specific type of structural defect was selected as an example and we expect the presence of different types of defects that are either native (ie. naturally occurring imperfections and growth-induced defects) or introduced deliberately to have a very similar effect on the ultimate strength of graphene monolayers prepared by mechanical exfoliation. The presence of these defects causes considerable disruption to the monolayer and as has been pointed out in the literature, this leads to the reduction of the strength of the monolayer

[40-42]. It can be seen from the cross-section of the monolayer that the presence of the defects leads to bulges in the graphene lattice.

Periodic boundary conditions were applied along both directions. Using molecular mechanical simulations, the whole sample was fully relaxed without strain, in which the interaction between carbon atoms is described by the adaptive intermolecular reactive empirical bond order (AIREBO) type empirical potential [43]. The tensile strains were applied by enlarging the lattice constant along the $Y$ direction. For each strained sample, the structure was relaxed statically. The strains of bonds in the central region are plotted in **Figure 9b** for an overall strain of about 2%. The calculations show that the strain is concentrated mainly in the C-C bonds in the heptagonal ring and decays quickly with bond position away from the heptagon (**Figure 9b**). In contrast, in the green circled areas, the bonds are actually subjected to a compressive strain. Thus, the whole area might be regarded as a *pseudo crack* and overall the presence of these two simple defect pairs can give rise to a stress concentration of the order of 6 in the graphene monolayer. This would ultimately reduce the strength of a graphene monolayer from 130 GPa to around 20 GPa.

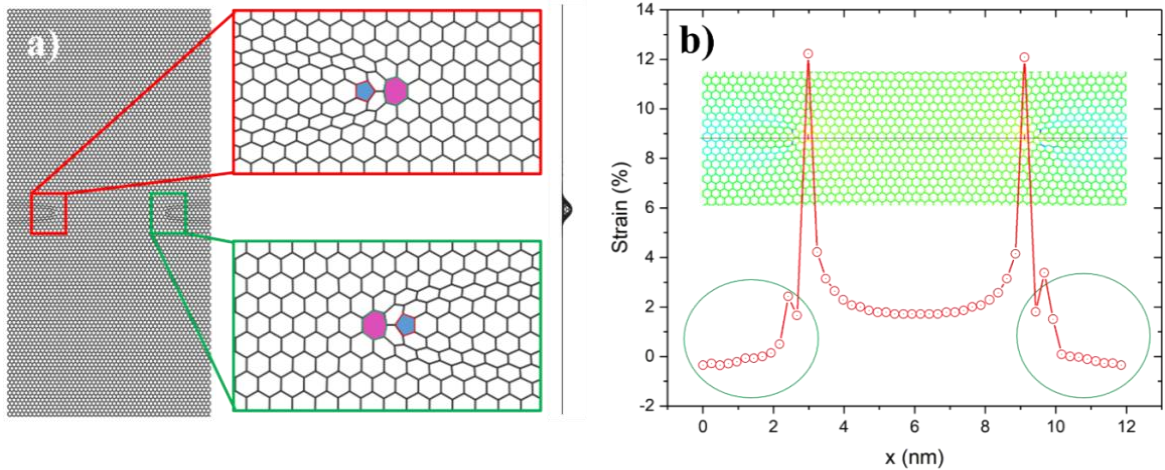

**Figure 9** a) Model of a pair of 7-5 defects of opposite sign in a graphene lattice and a cross section showing the bulges in the lattice. b) The variation of local bond strain along the central black line in the middle of the graphene monolayer from a) containing the pair of 7-5 defects. The overall applied strain is ~2% in the vertical direction.

### 3.6 Effect of flake size upon strength

The above analysis is consistent with the strength of flakes, being controlled by internal defects. Hence, when graphene is used in a nanocomposite, a balance must be kept in having flakes that are long enough to give efficient stress transfer [25] but are not so large that their strength is so low that they undergo premature fracture at low stress.

The failure strengths are plotted against the flake width again for all 13 graphene flakes in **Figure 10**. It was found that flakes with widths greater than 18 μm (dashed line) tended to fracture whereas for those narrower than this width, the interface between the graphene flake and PMMA substrate failed before fracture could occur. On the right side of the dashed line, it can be observed that the strength of graphene flakes decreases with increasing flake width. The lowest fracture strength shown in **Figure 10** is only around 3 GPa. This may also explain why in an example of model graphene nanocomposites [21], the derived fracture strength from tensile tests was found to be only around 10 GPa, again much lower than 130 GPa, as Lee *et al.* have reported [2].

To fit the experimental data in **Figure 10**, we considered the two mechanisms of failure of the graphene monolayers studied here; a) the fracture of the flake and b) the breakdown of the flake/polymer interface. In order to study the behaviour of the flakes that fractured as a result of the application of strain, we applied the Weibull theory, where from equation (8), the strength of the graphene flake as a function of width is given by:

$$\sigma_f = K \left(\frac{1}{w}\right)^{1/m} \tag{9}$$

Using the parameters from **Figure 8**, if we substitute the Weibull parameter $m = 1.37$ and set the constant $K = 71$ GPa for a flake with a width of 1 μm, we can see from

**Figure 10**, that the theoretical (red) line predicts quite well the strength of the flakes that fractured, indicating a strength lower than 10 GPa for a width greater than ~20 μm.

The breakdown of the flake/polymer interface is the mechanism of failure of the monolayer graphene flakes studied that were too narrow to fracture. If we assume that the breakdown of the interface occurs when a critical shear stress is reached at the end of the flake, the shear stress $\tau_i$ in the flake is given by equation (4), already presented in a previous section:

At the ends of the flake $x/l = 0.5$, so equation (4) becomes:

$$\tau_i = nE_g e_m \tanh\left(\frac{ns}{2}\right) \tag{10}$$

and if the interface fails at a critical value of $\tau_i$, then the breakdown stress is given by:

$$\sigma_b = E_g e_m = \frac{\tau_i}{n \tanh\left(\frac{nl}{2t}\right)} \tag{11}$$

This equation shows that the breakdown stress is a function of the flake length, $l$. If it is assumed to a first approximation that the flake width and length are similar (i.e. they are roughly square or circular) then this equation can also be plotted in **Figure 10**. After substituting using appropriate parameters ($\tau_i = 1.1$ MPa, $ns \sim 10$ and $l \sim w$), this equation produces the blue curve in the figure. It is important to note that the interfacial failure stress is predicted to fall to a plateau value, of the order of ~ 11 GPa, for flakes with widths greater than 20 μm, while for smaller widths, the predicted strength increases exponentially following the experimental data points for the flakes that undergo interfacial failure. The behaviour of the microribbons seems, however, to be an exception to this observation. They are too narrow to undergo fracture and their very high aspect ratios, $s$, means that $l >> w$ and they will undergo interfacial failure at the plateau value of around 11 GPa.

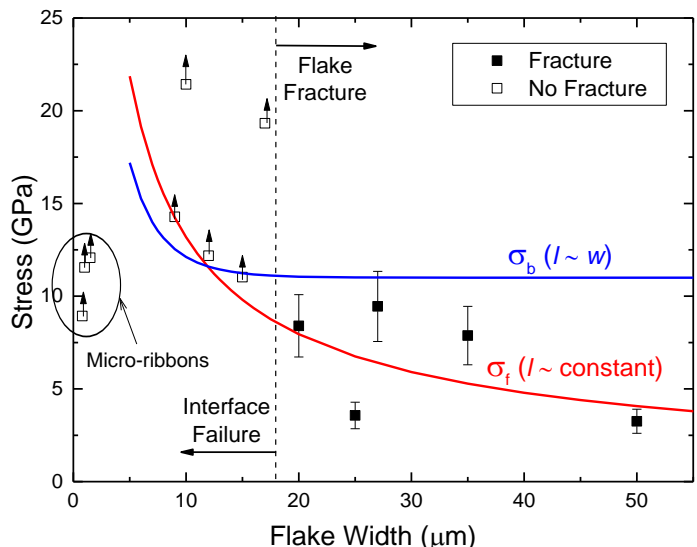

**Figure 10** Dependence of the graphene failure strength versus the flake width. The blue curve ($\sigma_b$) predicts the strength of the flakes where failure initiated by breakdown of the flake/polymer interface, while the red curve ($\sigma_f$) represents the strength of the flakes, where the main failure mechanism was the fracture of the flakes during strain.

The two theoretical curves cross near the dashed line that separates the flakes that fractured or did not fracture under strain. This implies that the failure mechanism will be either flake fracture or interface debonding failure, depending upon which one can occur at the lowest stress. The tendency for narrow flakes to undergo interfacial failure rather than fracture shows the limitations of the testing technique whereby stress is transferred from a PMMA beam to the flake through shear lag. If one is only interested in flake fracture then it would be better to grip the end of the flakes and apply the stress directly, which is much more difficult to do experimentally [44]. Nevertheless, the PMMA beam method is a realistic was to deform the graphene since in nanocomposites the flakes are deformed by shear stress transfer at the interface with the polymer matrix [8]. In addition, the tailoring of the flakes to identical lengths with different widths would be ideal in order to reach firm conclusions regarding the effect of the width of the flakes on the ultimate strength. Unfortunately, this is impossible to do by the tape exfoliation method and the application of other methods to achieve this may additionally alter the morphology of the flakes significantly. This present study has shown how in such circumstances, there will be a balance between the likelihood for a flake in the nanocomposite to either fracture or debond depending upon its size and shape. Stronger flake-matrix interfaces would be expected to lead to better-performing nanocomposites, as is found in practice [8].

## 4. Conclusions

In summary, this work presents a systematic investigation of the strength of monolayer graphene via the application of Raman spectroscopy and computational quantum mechanical modelling. It is shown that the strength of monolayer graphene is normally well below the 130 GPa measured by nanoidentation and reported previously [2]. It is found that that strength of flakes tends to decrease as they increase in size. The actual fracture strength of a material is invariably lower than its theoretical value because most finite-sized materials contain defects that concentrate stress. The molecular mechanical simulation results revealed that even if the monolayer graphene is crack-free, the presence of a pair of topological defects can cause considerable disruption in the monolayer, reducing its strength to around 20 GPa. Deformation of the flakes has been shown to occur through interfacial stress transfer from the PMMA substrate *via* a shear-lag process and relatively narrow monolayer flakes tend to undergo interfacial failure rather than fracture. The findings outlined above have added considerably to our understanding of the mechanical properties of mechanically-exfoliated graphene monolayers.

Overall, this present study clearly identifies that strain mapping *via* Raman spectroscopy can be successfully applied for the evaluation of the mechanical properties of finite-sized 2D flakes. The implications of this study in fields such as polymer composites reveal that even though long graphene flakes are needed for size of flakes should not be so large, otherwise they will undergo premature fracture at low stress.

## Acknowledgements

This project has received funding from the European Union's Horizon 2020 research and innovation programme under grant agreement No 785219.

## SUPPORTING INFORMATION

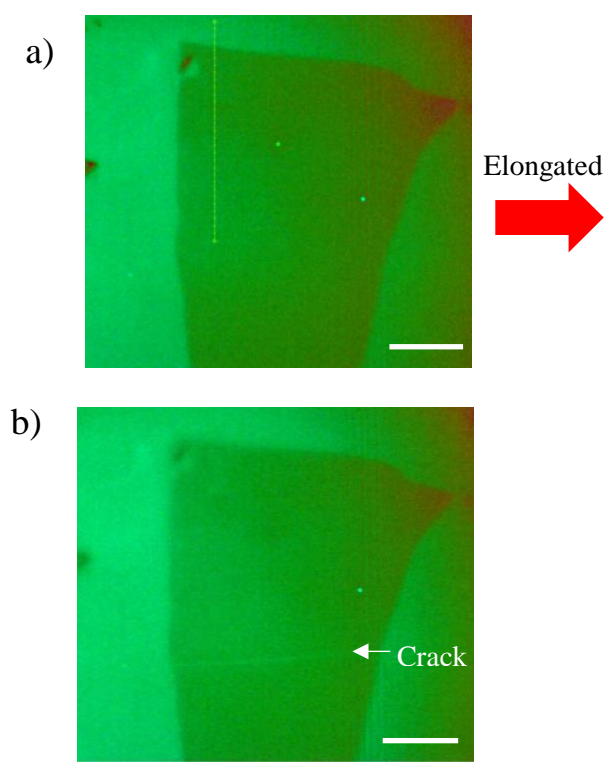

**Figure S1** a) and b) The optical images of Flake 2 before bending and after fracture captured in the colour palette mode.